\documentstyle[aps,prl,twocolumn]{revtex}

\begin{document}
\draft
\title{Quantum mechanical time-delay matrix in chaotic scattering}

\author{P. W. Brouwer,
  K. M. Frahm,\thanks{Present address: Laboratoire de Physique
  Quantique, UMR 5626 du CNRS,
  Universit\'e Paul Sabatier, 31062 Toulouse Cedex 4, France}
   and C. W. J. Beenakker}

\address{Instituut-Lorentz, Leiden University, P.O. Box 9506, 2300 RA
Leiden, The Netherlands
\medskip \\ \parbox{14cm}{\rm
We calculate the probability distribution of the matrix
$Q=-i\hbar\,S^{-1} \partial S/\partial E$ for a chaotic system
with scattering matrix $S$ at energy $E$.
The eigenvalues $\tau_j$ of $Q$ are the so-called proper delay
times, introduced by E.\ P.\ Wigner and F.\ T.\ Smith to
describe the time-dependence of a scattering process.
The distribution of the inverse delay times turns out to be
given by the Laguerre ensemble from random-matrix theory.
}}

\maketitle

\narrowtext

Eisenbud \cite{eisenbud} and Wigner \cite{wigner1} introduced the notion of
time delay in a quantum mechanical scattering problem. Wigner's one-dimensional
analysis was generalized to an $N \times N$ scattering matrix $S$ by Smith
\cite{Smith}, who studied the Hermitian energy derivative $Q = -i \hbar S^{-1}
\partial S/\partial E$ and interpreted its diagonal elements as the delay time
for a wave packet incident in one of the $N$ scattering channels. The matrix
$Q$ is called the Wigner-Smith time-delay matrix and its eigenvalues $\tau_1,
\tau_2, \ldots, \tau_N$ are called proper delay times.

Recently, interest in the time-delay problem was revived in the context of
chaotic scattering \cite{BlumelSmilansky1}. There is considerable theoretical
\cite{BlumelSmilansky1,BlumelSmilansky2,LW,brouwer2} and experimental
\cite{Smilansky,microwave1,microwave2} evidence that an ensemble of chaotic
billiards containing a small opening (through which $N$ modes can propagate at
energy $E$) has a uniform distribution of $S$ in the group of $N \times N$
unitary matrices --- restricted only by fundamental symmetries. This universal
distribution is the circular ensemble of random-matrix theory \cite{Mehta},
introduced by Dyson for its mathematical simplicity \cite{Dyson}. The
eigenvalues $e^{i \phi_n}$ of $S$ in the circular ensemble are distributed
according to
\begin{equation}
  P(\phi_1,\phi_2,\ldots,\phi_N) \propto \prod_{n < m} |e^{i \phi_n} - e^{i
\phi_m}|^{\beta}, \label{eq:circ}
\end{equation}
with the Dyson index $\beta=1$, $2$, $4$ depending on the presence or absence
of time-reversal and spin-rotation symmetry.

No formula of such generality is known for the time-delay matrix, although many
authors have worked on this problem
\cite{LW,lyub,Bauer,Harney,Eckhardt,Izrailev,Lehmann,FS1,SZZ,GMB,BB,MJP}. An
early result, $\langle \mbox{tr}\, Q \rangle = \tau_H$, is due to Lyuboshits
\cite{lyub}, who equated the ensemble average of the sum of the delay times
$\mbox{tr}\, Q = \sum_{n=1}^{N} \tau_n$ to the Heisenberg time $\tau_H = 2 \pi
\hbar/\Delta$ (with $\Delta$ the mean level spacing of the closed system). The
second moment of $\mbox{tr}\, Q$ was computed by Lehmann et al.\ \cite{Lehmann}
and by
Fyodorov and Sommers \cite{FS1}. The distribution of $Q$ itself is not known,
except for $N=1$ \cite{FS1,GMB}. The trace of $Q$ determines the density of
states \cite{Akkermans}, and is therefore sufficient for most thermodynamic
applications \cite{GMB}. For applications to quantum transport, however, the
distribution of all individual eigenvalues $\tau_n$ of $Q$ is needed, as well
as the distribution of the eigenvectors \cite{brouwer1}.

The solution of this 40 year old problem is presented here. We have found that
the eigenvalues of $Q$ are independent of $S$ \cite{foot3}. The distribution of
the inverse delay times $\gamma_n = 1/\tau_n$ turns out to be the Laguerre
ensemble of random-matrix theory,
\begin{equation}
  P(\gamma_1,\ldots,\gamma_N) \propto
  \prod_{i<j} |\gamma_{i} - \gamma_{j}|^{\beta}
  \prod_k \gamma_{k}^{\beta N/2}
  e^{-{\beta \tau_H \gamma_k/2}}, \label{eq:Lag}
\end{equation}
but with an unusual $N$-dependent exponent. (The function $P$ is zero if any
one of the $\tau_n$'s is negative.) The correlation functions of the $\tau_n$'s
consist of series over (generalized) Laguerre polynomials \cite{SlevinNagao},
hence the name ``Laguerre ensemble''. The eigenvectors of $Q$ are not
independent of $S$, unless $\beta=2$ (which is the case of broken time-reversal
symmetry). However, for any $\beta$ the correlations can be transformed away if
we replace $Q$ by the symmetrized matrix
\begin{equation}
  Q_E = -i \hbar\, S^{-1/2} {\partial S \over \partial E} S^{-1/2},
  \label{eq:QE}
\end{equation}
which has the same eigenvalues as $Q$. The matrix of eigenvectors $U$ which
diagonalizes $Q_E = U \mbox{diag($\tau_1,\ldots,\tau_N$)} U^{\dagger}$ is
independent of $S$ and the $\tau_n$'s, and uniformly distributed in the
orthogonal, unitary, or symplectic group (for $\beta=1$, $2$, or $4$,
respectively). The distribution (\ref{eq:Lag}) confirms the conjecture by
Fyodorov and Sommers \cite{FS1} that the distribution of $\mbox{tr}\, Q$ has an
algebraic tail $\propto (\mbox{tr}\, Q)^{-2-\beta N/2}$.

Although the time-delay matrix was interpreted by Smith as a representation of
the ``time operator'', this interpretation is ambiguous \cite{FS1}. The
ambiguity arises because a wavepacket has no well-defined energy. There is no
ambiguity in the application of $Q$ to transport problems where the incoming
wave can be regarded monochromatic, like the low-frequency response of a
chaotic cavity \cite{GMB,BB,BPT1993} or the Fermi-energy dependence of the
conductance \cite{brouwer1}. In the first problem, time delay is described by
complex reflection (or transmission) coefficients $R_{mn}(\omega)$,
\begin{mathletters} \label{eq:cap}
\begin{eqnarray}
  R_{mn}(\omega) &=&
    R_{mn}(0) [1 + i \omega \tau_{mn} + {\cal O}(\omega^2)], \\
  R_{mn}(0) &=& |S_{mn}|^2,\ \
    \tau_{mn} = \mbox{Im}\, \hbar S_{mn}^{-1} \partial S_{mn}/\partial E.
\end{eqnarray}
\end{mathletters}%
The delay time $\tau_{mn}$ determines the phase shift of the a.c.\ signal and
goes back to Eisenbud \cite{eisenbud}. With respect to a suitably chosen basis,
we may require that both the matrices $R_{mn}(0)$ and $\tau_{mn}$ are diagonal.
Then we have
\begin{eqnarray}
  R_{mn}(\omega) = \delta_{mn}[ 1 + i \omega \tau_{m} + {\cal O}(\omega^2)],
\end{eqnarray}
where the $\tau_{m}$ ($m=1,\ldots,N$) are the proper delay times (eigenvalues
of the Wigner-Smith time-delay matrix $Q$). For electronic systems, the ${\cal
O}(\omega)$ term of $R_{mn}(\omega)$ is the capacitance. Hence, in this
context, the proper delay times have the physical interpretation of
``capacitance eigenvalues'' \cite{footnote_on_selfconsistency}.

We now describe the derivation of our results. We start with some general
considerations about the invariance properties of the ensemble of
energy-dependent scattering matrices $S(E)$, following Wigner \cite{wigner2},
and Gopar, Mello, and B\"uttiker \cite{GMB}. The $N \times N$ matrix $S$ is
unitary for $\beta=2$ (broken time-reversal symmetry), unitary symmetric for
$\beta=1$ (unbroken time-reversal and spin-rotation symmetry), and unitary
self-dual for $\beta=4$ (unbroken time-reversal and broken spin-rotation
symmetry). The distribution functional $P[S(E)]$ of a chaotic system is assumed
to be invariant under a transformation
\begin{equation}
  S(E) \to V S(E) V', \label{eq:WignerConj}
\end{equation}
where $V$ and $V'$ are arbitrary unitary matrices which do not depend on $E$
($V' = V^{\rm T}$ for $\beta=1$, $V' = V^{\rm R}$ for $\beta=4$, where T
denotes the transpose and R the dual of a matrix). This invariance property is
manifest in the random-matrix model for the $E$-dependence of the scattering
matrix given in Ref.\ \onlinecite{BB}. A microscopic justification starting
from the Hamiltonian approach to chaotic scattering \cite{vwz} is given in
Ref.\ \onlinecite{unpublished}. Eq.\ (\ref{eq:WignerConj}) implies with $V = V'
= i S^{-1/2}$ that
\begin{equation}
  P(S,Q_E) = P(-\openone, Q_E). \label{eq:Sinv2}
\end{equation}
Here $P(S,Q_E)$ is the joint distribution of $S$ and $Q_E$, defined with
respect to the standard (flat) measure $dQ_E$ for the Hermitian matrix $Q_E$
and the invariant measure $dS$ for the unitary matrix $S$. From Eq.\
(\ref{eq:Sinv2}) we conclude that $S$ and $Q_E$ are statistically uncorrelated;
Their distribution is completely determined by its form at the special point
$S=-\openone$.

The distribution of $S$ and $Q_E$ at $S=-\openone$ is computed using
established methods of random-matrix theory \cite{Mehta,vwz}. The $N \times N$
scattering matrix $S$ is expressed in terms of the eigenvalues $E_{\alpha}$ and
the eigenfunctions $\psi_{n \alpha}$ of the $M \times M$ Hamiltonian matrix
${\cal H}$ of the closed chaotic cavity \cite{LW},
\begin{eqnarray}  \label{eq:SK}
  S &=& {1 - i K \over 1 + i K},\ \
  K_{mn} = {\Delta M \over \pi}
  \sum_{\alpha=1}^{M} {\psi_{m \alpha}^{\vphantom{*}} \psi_{n \alpha}^{*}
   \over E - E_{\alpha}}.
\end{eqnarray}
The Hermitian matrix ${\cal H}$ is taken from the Gaussian orthogonal (unitary,
symplectic) ensemble \cite{Mehta}, $P({\cal H}) \propto \exp(-\beta \pi^2
\mbox{tr}\, {\cal H}^2/4 \Delta^2 M)$. This implies that the eigenvector
elements $\psi_{j \alpha}$ are Gaussian distributed real (complex, quaternion)
numbers for $\beta=1$ ($2$, $4$), with zero mean and with variance $M^{-1}$,
and that the eigenvalues $E_{\alpha}$ have distribution
\begin{equation}
  P(\{E_{\alpha}\}) \propto \prod_{\mu < \nu} |E_{\mu} - E_{\nu}|^{\beta}
  \prod_{\mu} e^{-\beta \pi^2 E_{\mu}^2/4 \Delta^2 M}. \label{eq:Edistr}
\end{equation}
The limit $M \to \infty$ is taken at the end of the calculation.

The probability $P(-\openone,Q_E)$ is found by inspection of Eq.\ (\ref{eq:SK})
near $S=-\openone$. The case $S=-\openone$ is special, because $S$ equals
$-\openone$ only if the energy $E$ is an (at least) $N$-fold degenerate
eigenvalue of ${\cal H}$. For matrices $S$ in a small neighborhood of
$-\openone$, we may restrict the summation in Eq.\ (\ref{eq:SK}) to those $N$
energy levels $E_{\alpha}$, $\alpha=1,\ldots,N$, that are (almost) degenerate
with $E$ (i.e.\ $|E-E_{\alpha}| \ll \Delta$). The remaining $M - N$ eigenvalues
of ${\cal H}$ do not contribute to the scattering matrix. This enormous
reduction of the number of energy levels involved provides the simplification
that allows us to compute the complete distribution of the matrix $Q_E$.

We arrange the eigenvector elements $\psi_{n \alpha}$ into an $N \times N$
matrix $\Psi_{j \alpha} = \psi_{j \alpha} M^{1/2}$. Its distribution $P(\Psi)
\propto \exp(-\beta \mbox{tr}\, \Psi \Psi^{\dagger}/2)$ is invariant under a
transformation $\Psi \to \Psi O$, where $O$ is an orthogonal (unitary,
symplectic) matrix. We use this freedom to replace $\Psi$ by the product $\Psi
O$, and choose a uniform distribution for $O$. We finally define the $N \times
N$ Hermitian matrix $H_{ij} = \sum_{\alpha=1}^{N} O_{i\alpha} (E_{\alpha}-E)
O_{j\alpha}^{*}$. Since the distribution of the energy levels $E_{\alpha}$
close to $E$ is given by $\prod_{\mu < \nu} |E_{\mu} - E_{\nu}|^{\beta}$ [cf.\
Eq.\ (\ref{eq:Edistr})], it follows that the matrix $H$ has a uniform
distribution near $H = 0$. We then find
\begin{mathletters}
\begin{eqnarray}
  S   &=& -\openone + (i \tau_H/\hbar) \Psi^{\dagger-1} H \Psi^{-1},\\
  Q_E &=& \tau_H \Psi^{\dagger-1} \Psi^{-1}.
  \label{eq:QA}
\end{eqnarray}
\end{mathletters}
Hence the joint distribution of $S$ and $Q_E$ at $S=-\openone$ is
given by
\begin{eqnarray}
  P(-\openone,Q_E) &\propto& \int d\Psi \, d H \,
  e^{-\beta\, \mbox{tr}\, \Psi \Psi ^{\dagger}/2}
  \nonumber \\ && \mbox{} \times
  \delta(\Psi^{\dagger-1} H \Psi^{-1})\,
  \delta(Q_E - \tau_H \Psi^{\dagger-1} \Psi^{-1}) \nonumber \\ &=&
  \int d\Psi\,
  e^{-\beta\, \mbox{tr}\, \Psi \Psi ^{\dagger}/2}
  \left (\det\Psi \Psi^{\dagger} \right)^{(\beta N + 2 - \beta)/2}
  \nonumber \\ && \mbox{} \times
  \delta(Q_E - \tau_H \Psi^{\dagger-1} \Psi^{-1}). \label{eq:PSQQG}
\end{eqnarray}
The remaining integral depends entirely on the positive-definite Hermitian
matrix $\Gamma = \Psi \Psi^{\dagger}$. In Refs.\ \onlinecite{SlevinNagao} and
\onlinecite{BrezinHikamiZee} it is shown that
\begin{equation} \label{eq:LagG}
  \int d\Psi f(\Psi \Psi^{\dagger}) =
  \int d\Gamma \left( \det \Gamma \right)^{(\beta-2)/2} f(\Gamma)
\Theta(\Gamma),
\end{equation}
where $\Theta(\Gamma) =1$ if all eigenvalues of $\Gamma$ are positive and $0$
otherwise, and $f$ is an arbitrary function of $\Gamma = \Psi \Psi^{\dagger}$.
Integration of Eq.\ (\ref{eq:PSQQG}) with the help of Eq.\ (\ref{eq:LagG})
finally yields the distribution (\ref{eq:Lag}) for the inverse delay times and
the uniform distribution of the eigenvectors, as advertised.

In addition to the energy derivative of the scattering matrix, one may also
consider the derivative with respect to an external parameter $X$, such as the
shape of the system, or the magnetic field \cite{FS1,SZZ}. In random-matrix
theory, the parameter dependence of energy levels and wavefunctions is
described through a parameter dependent $M \times M$ Hermitian matrix ensemble,
\begin{equation}
  {\cal H}(X) = {\cal H} + M^{-1/2} X\, {\cal H}', \label{eq:XHam}
\end{equation}
where ${\cal H}$ and ${\cal H}'$ are taken from the same Gaussian ensemble. We
characterize $\partial S/\partial X$ through the symmetrized derivative
\begin{equation}
  Q_X = -i S^{-1/2} {\partial S \over \partial X} S^{-1/2},
\end{equation}
by analogy with the symmetrized time-delay matrix $Q_E$ in Eq.\ (\ref{eq:QE}).
To calculate the distribution of $Q_X$, we assume that the invariance
(\ref{eq:WignerConj}) also holds for the $X$-dependent ensemble of scattering
matrices. (A random-matrix model with this invariance property is given in
Ref.\ \onlinecite{BrouwerBeenakker1996a}). Then it is sufficient to consider
the special point $S=-\openone$. From Eqs.\ (\ref{eq:QA}) and (\ref{eq:XHam})
we find
\begin{equation}
  Q_X = \Psi^{\dagger-1} H' \Psi^{-1},\ \ P(H') \propto \exp(-\beta \mbox{tr}\,
H'^2/16),
\end{equation}
where $H'_{\mu \nu} = -(\tau_H/\hbar) M^{-1/2} \sum_{i,j} \psi_{i \mu}^{*}
{\cal H}'_{ij} \psi_{j \nu}^{\vphantom{*}}$. A calculation similar to that of
the distribution of the time-delay matrix shows that the distribution of $Q_X$
is a Gaussian, with a width set by $Q_E$,
\begin{eqnarray}
  && P(S,Q_E,Q_X) \propto (\det Q_E)^{-2\beta N-3+3\beta/2}\,
  \nonumber \\ && \ \ \ \mbox{} \times
  \exp\left[{-{\beta \over 2}\,\mbox{tr}\,
    \left(\tau_H Q_E^{-1} +
    {1 \over 8}(\tau_H Q_E^{-1} Q_X^{\vphantom{1}})^2\right)}\right].
    \label{eq:PSQQ}
\end{eqnarray}
The fact that delay times set the scale for the sensitivity to an external
perturbation in an open system is well understood in terms of classical
trajectories \cite{jalabert}, in the semiclassical limit $N \to \infty$. Eq.\
(\ref{eq:PSQQ}) makes this precise in the fully quantum-mechanical regime of a
finite number of channels $N$. Correlations between parameter dependence and
delay time were also obtained in Refs.\ \onlinecite{FS1,SZZ}, for the phase
shift derivatives $\partial \phi_j/\partial X$.

In summary, we have calculated the distribution of the Wigner-Smith time-delay
matrix for chaotic scattering. This is relevant for experiments on
frequency and parameter-dependent transmission through chaotic microwave
cavities \cite{microwave1,microwave2} or semiconductor quantum dots with
ballistic point contacts \cite{Westervelt}. The distribution (\ref{eq:circ})
has been known since Dyson's 1962 paper as the circular ensemble \cite{Dyson}.
It is remarkable that the Laguerre ensemble (\ref{eq:Lag}) for the (inverse)
delay times was not discovered earlier.

We acknowledge support by the Dutch Science Foundation NWO/FOM.


\begin{references}
\bibitem{foot3} The absence of correlations between the $\tau_n$'s and
        the $\phi_n$'s is a special property of the proper delay times.
        In contrast, the derivatives $\partial\phi_n/\partial E$
        considered in Refs.\ \onlinecite{FS1,SZZ} are correlated with
        the $\phi_n$'s.
\end{references}
\end{document}